\documentclass[journal]{IEEEtran}
\ifCLASSINFOpdf
\else
   \usepackage{graphicx}
   \graphicspath{{../eps/}}
   \DeclareGraphicsExtensions{.eps}
\fi

\usepackage{amsmath}
\usepackage{amsfonts,amssymb}
\usepackage{amssymb}
\usepackage{wrapfig}
\usepackage{psfrag}
\usepackage{epstopdf}
\usepackage{cite}
\usepackage{graphicx}
\usepackage{subfigure}
\usepackage{threeparttable}
\usepackage{cases}
\usepackage{subeqnarray}
\usepackage{color}
\usepackage{underscore}
\usepackage{verbatim}
\usepackage{bm}
\usepackage{stfloats}
\newtheorem{theorem}{Theorem}

\usepackage{algorithm}
\usepackage{algorithmic}

\begin{document}

\title{Delay Alignment Modulation: Enabling Equalization-Free Single-Carrier Communication}
%
%
%
\author{
        Haiquan~Lu
        and
        Yong~Zeng,~\IEEEmembership{Member,~IEEE}
\thanks{This work was supported by the National Key R\&D Program of China with Grant number 2019YFB1803400.}
\thanks{H. Lu and Y. Zeng are with the National Mobile Communications Research Laboratory, Southeast University, Nanjing 210096, China. Y. Zeng is also with the Purple Mountain Laboratories, Nanjing 211111, China (e-mail: \{haiquanlu, yong_zeng\}@seu.edu.cn). (\emph{Corresponding author: Yong Zeng.})}
\vspace{-2.5ex}
}

\maketitle

\begin{abstract}
 This paper proposes a novel broadband transmission technology, termed \emph{delay alignment modulation} (DAM), which enables the low-complexity equalization-free single-carrier communication, yet without suffering from inter-symbol interference (ISI). The key idea of DAM is to deliberately introduce appropriate delays for information-bearing symbols at the transmitter side, so that after propagating over the time-dispersive channel, all multi-path signal components will arrive at the receiver simultaneously and constructively. We first show that by applying DAM for the basic multiple-input single-output (MISO) communication system, an ISI-free additive white Gaussian noise (AWGN) system can be obtained with the simple zero-forcing (ZF) beamforming. Furthermore, the more general DAM scheme is studied with the ISI-maximal-ratio transmission (MRT) and the ISI-minimum mean-square error (MMSE) beamforming. Simulation results are provided to show that when the channel is sparse and/or the antenna dimension is large, DAM not only resolves the notorious practical issues suffered by orthogonal frequency-division multiplexing (OFDM) such as high peak-to-average-power ratio (PAPR), severe out-of-band (OOB) emission, and vulnerability to carrier frequency offset (CFO), with low complexity, but also achieves higher spectral efficiency due to the saving of guard interval overhead.
\end{abstract}


\IEEEpeerreviewmaketitle
\vspace{-0.5cm}
\section{Introduction}
Since its introduction around 1960s, orthogonal frequency-division multiplexing (OFDM) has been gradually evolved as the dominant transmission technology for broadband communication, with successful applications including the fourth- (4G) and fifth-generation (5G) cellular networks, and wireless local area network (WLAN) \cite{heath2018foundations}. As a digital multi-carrier technology, OFDM 
enables high-rate communications while circumventing the detrimental inter-symbol interference (ISI) \cite{heath2018foundations,goldsmith2005wireless}. In addition, adaptive modulation and multiple access can be flexibly applied for OFDM. However, it is also well known that OFDM suffers from some critical drawbacks, including the high peak-to-average-power ratio (PAPR) \cite{han2005overview}, the severe out-of-band (OOB) emission \cite{farhang2011ofdm}, as well as the vulnerability to carrier frequency offset (CFO) \cite{sathananthan2001probability}.

During the past few decades, numerous efforts have been devoted to addressing the above issues for OFDM. For instance, the simplest method for PAPR reduction is amplitude clipping \cite{han2005overview}. As an alternative of OFDM without suffering from the high PAPR issue, single-carrier frequency-division equalization (SC-FDE) has been adopted by 4G and 5G uplink communications. Furthermore, to reduce the OOB emission of OFDM, the windowing and filtering approach can be applied \cite{nissel2017filter}. Moreover, filter bank multi-carrier (FBMC) has been extensively studied as an alternative to OFDM with reduced OOB emission
\cite{nissel2017filter,farhang2011ofdm}. More recently, to deal with the severe CFO issue in high-mobility scenarios that are exacerbated at high carrier frequencies such as millimeter wave (mmWave) and Terahertz communications, a new multi-carrier technique termed \emph{orthogonal time frequency space (OTFS) modulation} was proposed \cite{hadani2017orthogonal}. However, such existing techniques for OFDM and its variants either incur performance loss or require complicated signal processing.

In this paper, by exploiting the abundant spatial dimension brought by large antenna arrays \cite{bjornson2019massive,lu2021communicating} and the multi-path sparsity of mmWave and Terahertz channels \cite{akdeniz2014millimeter,zeng2016millimeter,han2018propagation}, we propose a novel broadband transmission technology, termed \emph{delay alignment modulation} (DAM). DAM enables low-complexity equalization-free single-carrier communication, yet without suffering from ISI even in frequency-selective channels. The key idea of DAM is to deliberately introduce appropriate delays for the information-bearing symbols at the transmitter side, so that after propagating over the time-dispersive channel, all multi-path signal components will arrive at the receiver simultaneously and constructively.
Note that DAM is most effective for systems with multi-path sparsity and high spatial dimensions, which is expected to be the case for 6G mmWave/Terahertz communication with massive/extremely large antenna arrays. This is because when the spatial design degree of freedom (DoF) is much larger than the number of temporal-resolvable multi-paths, the propagation delay of each path can be pre-compensated independently without affecting all other paths, which makes it possible to perfectly align all multi-path signal components at the receiver.

In this paper, the proposed DAM is studied in details for the basic multiple-input single-output (MISO) communication system. We first show that with DAM, an ISI-free additive white Gaussian noise (AWGN) system can be obtained with the simple zero-forcing (ZF) beamforming. In addition, for the more general case when the ISI-ZF condition is infeasible due to the insufficient spatial dimensions, or when certain residual ISI is tolerable, the ISI-maximal-ratio transmission (MRT) and the ISI-minimum mean-square error (MMSE) beamforming design are studied. It is also worthwhile to remark that the idea of path delay pre-compensation at the transmitter was firstly proposed in our previous work \cite{zeng2016millimeter} and \cite{zeng2018multi}, but only for lens multiple-input
multiple-output (MIMO) systems.

Note that compared to OFDM and its different variants mentioned above, the proposed DAM technique has several appealing advantages. Firstly, it is a single-carrier broadband transmission technique, which inherently avoids those notorious drawbacks of OFDM (or more generally multi-carrier transmission), including the high PAPR, the severe OOB emission, and the vulnerability to CFO. Secondly, as elaborated later in this paper, DAM can achieve even higher spectral efficiency, since it requires a guard interval only for every channel coherence block, instead of for every symbol as for OFDM. Last but not least, DAM can be implemented with even lower complexity than OFDM (let alone than FBMC and OTFS), since delay pre-compensation at the transmitter simply means sequence shift of the information-bearing symbols. Despite of such promising advantages, we do not intend to claim that DAM can be used to replace OFDM, since it gives the best performance only for multi-path sparse channels. Besides, DAM also faces its own challenges, such as its critical dependence on the channel state information and the non-trivial extension to multiple access scenarios, which deserve further studies.

\vspace{-0.5cm}
\section{System Model and Delay Alignment Modulation}
 \begin{figure}[!t]
 \setlength{\abovecaptionskip}{-0.15cm}
 \setlength{\belowcaptionskip}{-0.3cm}
 \centering
 \centerline{\includegraphics[width=3.2in,height=1.9in]{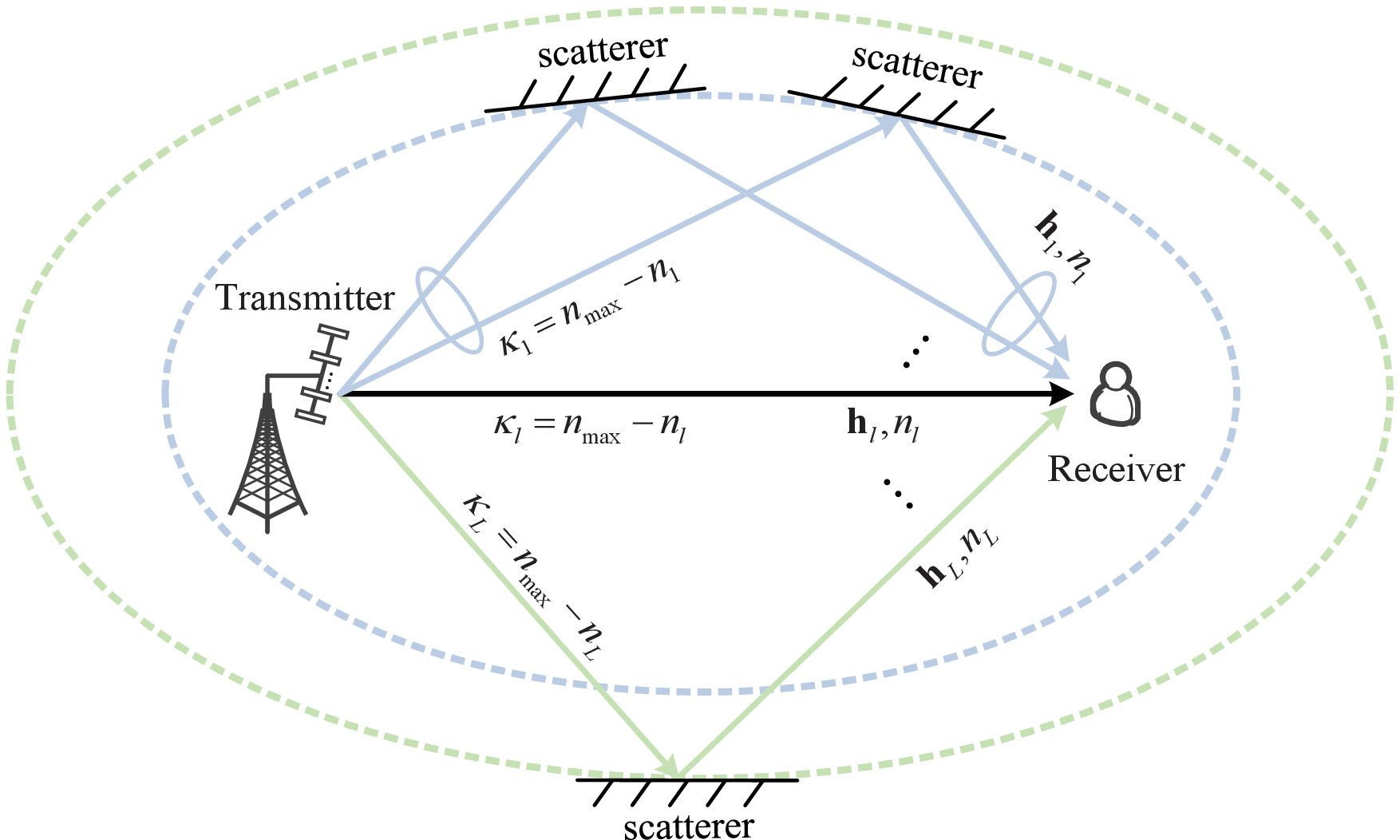}}
 \caption{A MISO communication system with delay alignment modulation.}
 \label{systemModel}
 \vspace{-0.7cm}
 \end{figure}
 We consider a MISO communication system, as shown in  Fig.~\ref{systemModel}, where the transmitter is equipped with $M$ antennas and the receiver has one antenna. The available bandwidth is $B$. Under the multi-path environment, the discrete-time equivalent of the channel impulse response can be expressed as
 \begin{equation}\label{channelImpulseResponse}
 {{\bf{h}}^H}\left[ n \right] = \sum\nolimits_{l = 1}^L {{\bf{h}}_l^H\delta \left[ {n - {n_l}} \right]},
 \end{equation}
 where $L$ is the number of temporal-resolvable multi-paths with delay resolution ${1}/{B}$, ${\bf{h}}_l \in {{\mathbb{C}}^{M \times 1}}$ denotes the channel vector for the $l$th multi-path, and ${n_l}$ denotes its delay in terms of symbol durations. For multi-path sparsity channels in mmWave and Teraherta communications with large antenna arrays, the number of multi-paths is usually much smaller than that of antennas, i.e., $L \ll M$ \cite{akdeniz2014millimeter,zeng2016millimeter,han2018propagation}. Note that the multi-path in \eqref{channelImpulseResponse} is defined in the temporal domain based on the resolvable delays, while each multi-path may include several sub-paths that have a common delay but different angle of departures (AoDs), as illustrated in Fig.~\ref{systemModel}. As a result, ${\bf{h}}_l$ in \eqref{channelImpulseResponse} can be in general modelled as  ${{\bf{h}}_l} = {\alpha _l}\sum\nolimits_{i = 1}^{{\mu _l}} {{\upsilon _{li}}{\bf{a}}\left( {{\theta _{li}}} \right)}$, where ${\alpha _l}$ and $\mu_l$ denote the complex-valued path gain and the number of sub-paths for the $l$th multi-path, respectively, ${\upsilon _{li}} = \frac{1}{{\sqrt {{\mu _l}} }}{e^{j{\phi _{li}}}}$ denotes the complex coefficient of the $i$th sub-path of path $l$, and $\theta_{li}$ and ${{\bf{a}}\left( {\theta _{li}} \right)} \in {{\mathbb{C}}^{M \times 1}}$ denote the AoD and the transmit array response vector of the $i$th sub-path of path $l$, respectively.

 Denote by ${\bf{x}}\left[ n \right] \in {{\mathbb{C}}^{M \times 1}}$ the discrete-time equivalent of the transmitted signal. Then the received signal is
 \begin{equation}\label{generalReceivedSignal}
 \hspace{-0.7ex}
 \begin{aligned}
 y\left[ n \right] {\rm =} {{\bf{h}}^H}\left[ n \right]*{\bf{x}}\left[ n \right] + z\left[ n \right] {\rm =} \sum\nolimits_{l = 1}^L {{\bf{h}}_l^H{\bf{x}}\left[ {n - {n_l}} \right]}  + z\left[ n \right],
 \end{aligned}
 \end{equation}
 where $z\left[ n \right] \sim {\cal C}{\cal N}\left( {0,{\sigma ^2}} \right)$ is the AWGN. Denote by ${n_{\min }} \buildrel \Delta \over = \mathop {\min }\limits_{1 \le l \le L} {n_l}$ and ${n_{\max }} \buildrel \Delta \over = \mathop {\max }\limits_{1 \le l \le L} {n_l}$ the minimum and maximum delay over all the $L$ multi-paths, respectively, and the channel delay spread is ${n_{\rm{span}}} = {n_{\max }} - {n_{\min }}$.

 Let $s\left[ n \right]$ be the independent and identically distributed (i.i.d.) information-bearing symbols with normalized power ${\mathbb{E}}{\rm{[|}}s[n]{{\rm{|}}^2}] = 1$. If the conventional signal-carrier transmission is applied with ${\bf{f}} \in {{\mathbb{C}}^{M \times 1}}$ being the transmit beamforming vector, the transmitted signal can be written as ${\bf{x}}\left[ n \right] = {\bf{f}}s\left[ n \right]$. By substituting ${\bf{x}}\left[ n \right]$ into \eqref{generalReceivedSignal}, we have
 \begin{equation}\label{SCModulationReceivedSignal}
 \begin{aligned}
 y\left[ n \right] = \underbrace {{\bf{h}}_1^H{\bf{f}}s\left[ {n - {n_1}} \right]}_{{\rm{desired}}\;{\rm{signal}}} + \underbrace {\sum\nolimits_{l \ne 1}^L {{\bf{h}}_l^H{\bf{f}}s\left[ {n - {n_l}} \right]} }_{{\rm{ISI}}} + z\left[ n \right],
 \end{aligned}
 \end{equation}
 where we assume that the receiver is synchronized to path $1$. In this case, all the remaining $L-1$ multi-paths cause the detrimental ISI. Various techniques have been proposed to address the ISI problem, including time- and frequency-domain equalization, spread spectrum, and OFDM.

  \begin{figure}[!t]
  \setlength{\abovecaptionskip}{-0.1cm}
  \setlength{\belowcaptionskip}{-0.3cm}
  \centering
  \centerline{\includegraphics[width=3.0in,height=1.95in]{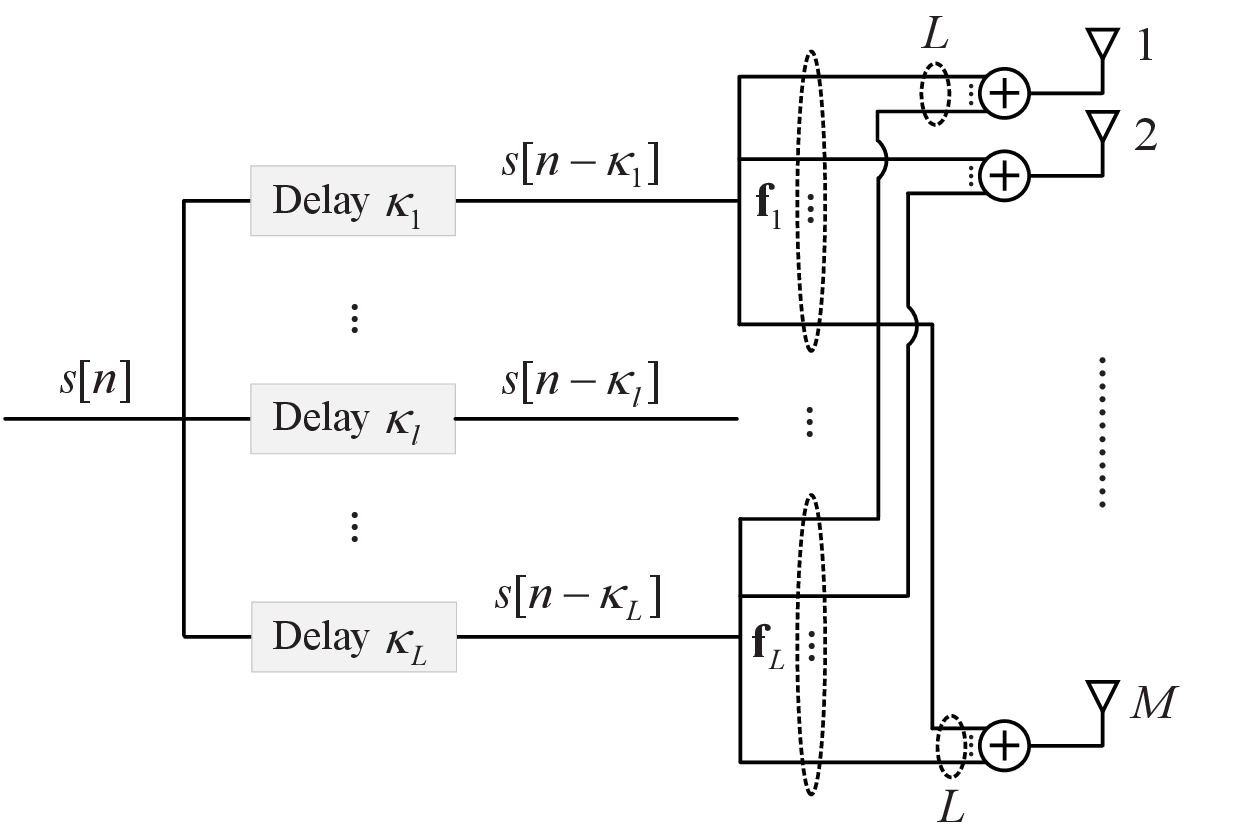}}
  \caption{Transmitter architecture for delay alignment modulation.}
  \label{DAMBlockDiagram}
  \vspace{-0.6cm}
  \end{figure}
 In this paper, DAM is proposed as a novel technique to enable low-complexity and equalization-free communication to address the ISI issue. With DAM, the transmitted signal is
 \begin{equation}\label{DAMTransmitSignal}
 {\bf{x}}\left[ n \right] = \sum\nolimits_{l = 1}^L {{{\bf{f}}_l}s\left[ {n - {\kappa _l}} \right]},
 \end{equation}
 where ${{{\bf{f}}_l}} \in {{\mathbb{C}}^{M \times 1}}$ denotes the transmit beamforming vector associated with path $l$, and $\kappa _l$ is the deliberately introduced delay for the symbol sequence $s\left[ n \right]$, with ${\kappa _l} \ne {\kappa _{l'}}$, $\forall l \ne l'$. The transmitter architecture of DAM is illustrated in Fig.~\ref{DAMBlockDiagram}, which can be easily implemented since delay compensation by $\kappa_l$ simply means time shift of the sequence $s\left[ n \right]$. The transmit power of ${\bf{x}}\left[ n \right]$ in \eqref{DAMTransmitSignal} is
 \begin{equation}\label{transmitPowerConstraint}
 \hspace{-0.75ex}
 \begin{aligned}
 {\mathbb{E}}\left[ {{{\left\| {{\bf{x}}\left[ n \right]} \right\|}^2}} \right] \mathop  = \limits^{\left( a \right)} \sum\nolimits_{l = 1}^L {{\mathbb{E}}\left[ {{{\left\| {{{\bf{f}}_l}s\left[ {n - {\kappa_l}} \right]} \right\|}^2}} \right]}  {\rm =} \sum\nolimits_{l = 1}^L {{{\left\| {{{\bf{f}}_l}} \right\|}^2}}  \le P,
 \end{aligned}
 \end{equation}
 where $P$ denotes the available transmit power, and ${\left( a \right)}$ holds since $s\left[ n \right]$ are independent across different $n$ and ${\kappa _l} \ne {\kappa _{l'}}$, $\forall l\neq l'$. By substituting \eqref{DAMTransmitSignal} into \eqref{generalReceivedSignal}, the received signal for DAM is
 \begin{equation}\label{DAMReceivedSignal}
 \begin{aligned}
 y\left[ n \right] &= \sum\nolimits_{l = 1}^L {{\bf{h}}_l^H{{\bf{f}}_l}s\left[ {n - {\kappa _l} - {n_l}} \right]}  + \\
 & \sum\nolimits_{l = 1}^L {\sum\nolimits_{l' \ne l}^L {{\bf{h}}_l^H{{\bf{f}}_{l'}}s\left[ {n - {\kappa _{l'}} - {n_l}} \right]} }  + z\left[ n \right].
 \end{aligned}
 \end{equation}
 It is observed from \eqref{DAMReceivedSignal} that the introduced delay $\kappa_l$, $\forall l$ impacts all the $L$ multi-path signal components. By letting ${\kappa _l} = {n_{\max }} - {n_l} \ge 0$, $\forall l$, we have
 \begin{equation}\label{DelayCompensationReceivedSignal}
 \begin{aligned}
 &y\left[ n \right]  = \Big( {\sum\nolimits_{l = 1}^L {{\bf{h}}_l^H{{\bf{f}}_l}} } \Big)s\left[ {n - {n_{\max }}} \right]+ \\
 & \sum\nolimits_{l = 1}^L {\sum\nolimits_{l' \ne l}^L {{\bf{h}}_l^H{{\bf{f}}_{l'}}s\left[ {n - {n_{\max }} + {n_{l'}} - {n_l}} \right]} }  + z\left[ n \right].
 \end{aligned}
 \end{equation}
 If the receiver is synchronized to the delay $n_{\max}$, then the first term in \eqref{DelayCompensationReceivedSignal} contributes to the desired signal, while the second term is the ISI. It is observed that the ISI cannot be completely eliminated by delay compensation alone. However, if $\{ {{\bf{f}}_l}\} _{l = 1}^L$ is designed so that
 \begin{equation}\label{DAMCondition}
 {\bf{h}}_l^H{{\bf{f}}_{l'}} = 0,\ \forall l \ne l',
 \end{equation}
 then the received signal in \eqref{DelayCompensationReceivedSignal} reduces to
 \begin{equation}\label{DelayCompensationReceivedSignalEquivalent}
 y\left[ n \right] = \Big( {\sum\nolimits_{l = 1}^L {{\bf{h}}_l^H{{\bf{f}}_l}} } \Big)s\left[ {n - {n_{\max }}} \right] + z\left[ n \right].
 \end{equation}
 It is observed from \eqref{DelayCompensationReceivedSignalEquivalent} that the received signal is simply the symbol sequence $s\left[ n \right]$ delayed by one single delay $n_{\max}$ with a multiplicative gain contributed by all the $L$ multi-paths. In other words, with the simple DAM in \eqref{DAMTransmitSignal} and the ISI-ZF beamforming in \eqref{DAMCondition}, the original multi-path frequency-selective channel has been transformed to the simple ISI-free AWGN channel with a single delay $n_{\max}$. Note that the ISI-ZF condition in \eqref{DAMCondition} can be satisfied as long as $M \ge L$, i.e., when the channel is sparse and/or the antenna dimension is large. On the other hand, when $M < L$, the perfect ISI-ZF beamforming is infeasible. In this case, the low-complexity ISI-MRT and the optimal ISI-MMSE beamforming can be applied for DAM based on the input-output relationship in \eqref{DelayCompensationReceivedSignal}, which are elaborated in the following.

\vspace{-0.3cm}
\section{ISI Beamforming and Performance Analysis}\label{sectionBeamformingOptimizationandPerformanceAnalysis}

\subsection{DAM with ISI-ZF Beamforming}
In this subsection, we aim to maximize the received signal-to-noise ratio (SNR) of DAM subject to the ISI-ZF condition \eqref{DAMCondition} via optimizing the beamforming vectors $\left\{ {{{\bf{f}}_l}} \right\}_{l = 1}^L$. The received SNR of \eqref{DelayCompensationReceivedSignalEquivalent} is
 \begin{equation}\label{DAMReceivedSNR}
 \gamma  = \frac{{{{\Big| {\sum\nolimits_{l = 1}^L {{\bf{h}}_l^H{{\bf{f}}_l}} } \Big|}^2}}}{{{\sigma ^2}}}.
 \end{equation}
 By discarding those constant terms, the beamforming problem can be formulated as
 \begin{equation}\label{originalProblem}
 \begin{aligned}
 \left( {{\rm{P1}}} \right)\ \ \mathop {\max }\limits_{\left\{ {{{\bf{f}}_l}} \right\}_{l = 1}^L} &\ \ {\Big| {\sum\nolimits_{l = 1}^L {{\bf{h}}_l^H{{\bf{f}}_l}} } \Big|^2}\\
 {\rm{s.t.}}&\ \ {\bf{h}}_l^H{{\bf{f}}_{l'}} = 0,\ \forall l \ne l',\ \ \sum\nolimits_{l = 1}^L {{{\left\| {{{\bf{f}}_l}} \right\|}^2}}  \le P. \nonumber
 \end{aligned}
 \end{equation}

 Let ${{\bf{H}}_l} = \left[ {{{\bf{h}}_1}, \cdots ,{{\bf{h}}_{l - 1}},{{\bf{h}}_{l + 1}}, \cdots ,{{\bf{h}}_L}} \right]$, $\forall l$. The ISI-ZF constraint of (P1) can be equivalently expressed as ${\bf{H}}_l^H{{\bf{f}}_l} = {{\bf{0}}_{\left( {L - 1} \right) \times 1}}$, $\forall l$. In other words, ${{\bf{f}}_l}$ should lie in the null space of ${\bf{H}}_l^H$. Further denote by ${{\bf{Q}}_l} \buildrel \Delta \over = {{\bf{I}}_M} - {{\bf{H}}_l}{\left( {{\bf{H}}_l^H{{\bf{H}}_l}} \right)^{ - 1}}{\bf{H}}_l^H$ the projection matrix into the space orthogonal to the columns of ${{\bf{H}}_l}$ \cite{brown2012practical}. Then we should have ${{\bf{f}}_l} = {{\bf{Q}}_l}{{\bf{b}}_l}$, $\forall l$, where ${{\bf{b}}_l} \in {{\mathbb{C}}^{M \times 1}}$ denotes the new vector to be designed. As a result, (P1) is equivalent to
 \begin{equation}\label{EquivalentProblem}
 \begin{aligned}
 \left( {{\rm{P2}}} \right)\ \ \mathop {\max }\limits_{\left\{ {{{\bf{b}}_l}} \right\}_{l = 1}^L} &\ \ {\Big| {\sum\nolimits_{l = 1}^L {{\bf{h}}_l^H{{\bf{Q}}_l}{{\bf{b}}_l}} } \Big|^2}\\
 {\rm{s.t.}}&\ \ \sum\nolimits_{l = 1}^L {{{\left\| {{{\bf{Q}}_l}{{\bf{b}}_l}} \right\|}^2}}  \le P. \nonumber
 \end{aligned}
 \end{equation}

 \begin{theorem}\label{DAMOptimalReceivedSNR}
  For DAM, the optimal ISI-ZF beamforming is
 \begin{equation}\label{optimalTransmitBeamforming}
 {\bf{f}}_l^{\rm{ZF}} = \frac{{\sqrt P {{\bf{Q}}_l}{{\bf{h}}_l}}}{{\sqrt {\sum\nolimits_{l = 1}^L {{{\left\| {{{\bf{Q}}_l}{{\bf{h}}_l}} \right\|}^2}} } }}, \ \forall l,
 \end{equation}
 and the resulting SNR is
 \begin{equation}\label{DAMReceivedOptimalSNRExpression}
 {\gamma _{{\rm{ZF}}}} = \bar P\sum\nolimits_{l = 1}^L {{{\left\| {{{\bf{Q}}_l}{{\bf{h}}_l}} \right\|}^2}},
 \end{equation}
  where $\bar P \buildrel \Delta \over = P/{\sigma ^2}$.
\end{theorem}

 \begin{IEEEproof}
 This result can be obtained according to the Cauchy-Schwarz inequality, which is omitted for brevity.
 \end{IEEEproof}

 Theorem \ref{DAMOptimalReceivedSNR} shows that by applying the ISI-ZF transmit beamforming for DAM so that the ISI is perfectly eliminated, all multi-path signal components are beneficial for the SNR enhancement, instead of being detrimental. This thus enables a new low-complexity single-carrier communication without having to perform complicated equalization nor  sophisticated multi-carrier transmission.

 \begin{theorem}\label{DAMReceivedSNRMRT}
 When the channel is sparse and/or the antenna dimension is large so that $M \gg L$, the optimal ISI-ZF beamforming and the received SNR respectively reduce to
 \begin{equation}\label{optimalMRTBeamforming}
 {\bf{f}}_l^{{\rm{ZF}}} \to  {\bf{f}}_{l}^{{\rm{MRT}}} = \frac{{\sqrt P {{\bf{h}}_l}}}{{\sqrt {\sum\nolimits_{l = 1}^L {{{\left\| {{{\bf{h}}_l}} \right\|}^2}} } }}, \ \forall l,
 \end{equation}
 \begin{equation}\label{DAMReceivedSNRMRTExpression}
 {\gamma _{{\rm{ZF}}}} \to \bar P\sum\nolimits_{l = 1}^L {{{\left\| {{{\bf{h}}_l}} \right\|}^2}}.
 \end{equation}
 \end{theorem}

 \begin{IEEEproof}
 This result can be obtained due to the fact $\frac{1}{M}{\bf{H}}_l^H{{\bf{h}}_l}\mathop  \to \limits^{\rm {a.s.}} {\bf{0}}$, as $M \to \infty$, with $\mathop  \to \limits^{\rm{a.s.}} $ denoting the almost sure convergence \cite{ngo2013energy}, which is omitted for brevity.
 \end{IEEEproof}

 Theorem \ref{DAMReceivedSNRMRT} is quite appealing for 6G communications with large antenna  arrays and mmWave/Terahertz communications, for which the low-complexity ISI-MRT beamforming approaches the performance of the ISI-ZF beamforming.

\vspace{-0.2cm}
\subsection{DAM with ISI-MRT and ISI-MMSE Beamforming}
 In this subsection, we study the more general case when the ISI-ZF condition in \eqref{DAMCondition} is infeasible, or when the beamforming design does not aim to completely remove the ISI, but rather tolerate some residual ISI.

 Let ${\cal L} \buildrel \Delta \over = \left\{ {l:l = 1, \cdots ,L} \right\}$ be the set of all multi-paths, and ${{\cal L}_l} \buildrel \Delta \over = {\cal L}\backslash l$ includes all other multi-paths excluding path $l$. Denote by ${\Delta _{l',l}} \buildrel \Delta \over = {n_{l'}} - {n_l}$ the \emph{delay difference} between path $l'$ and $l$. Then $\forall l \ne l'$, ${\Delta _{l',l}} \in \left\{ { \pm 1, \cdots , \pm {n_{\rm{span}}}} \right\}$. The received signal in \eqref{DelayCompensationReceivedSignal} can be written as
 \begin{equation}\label{delayDifferenceCompensationReceivedSignal}
 \begin{aligned}
 y\left[ n \right] & = \Big( {\sum\nolimits_{l = 1}^L {{\bf{h}}_l^H{{\bf{f}}_l}} } \Big)s\left[ {n - {n_{\max }}} \right] +  \\
 &\sum\nolimits_{l = 1}^L {\sum\nolimits_{l' \ne l}^L {{\bf{h}}_l^H{{\bf{f}}_{l'}}s\left[ {n - {n_{\max }} + {\Delta _{l',l}}} \right]} }  + z\left[ n \right].
 \end{aligned}
 \end{equation}
 Note that $\forall l \ne l'$, we have ${{\Delta _{l',l}}} \ne 0$. In order to derive the  signal-to-interference-plus-noise ratio (SINR) of \eqref{delayDifferenceCompensationReceivedSignal} with residual ISI, \eqref{delayDifferenceCompensationReceivedSignal} needs to be reformulated by grouping those interfering symbols with identical delay difference since they correspond to identical symbols \cite{zeng2018multi}. To this end, for each delay difference $i \in \left\{ { \pm 1, \cdots , \pm {n_{\rm{span}}}} \right\}$, define the following effective channel
 \begin{equation}\label{delayDifferenceChannel}
 {\bf{g}}_{l'}^H\left[ i \right] \buildrel \Delta \over = \left\{ \begin{split}
 &{\bf{h}}_l^H,\ \ {\rm{if}}\ \exists l \in {{\cal L}_{l'}},\ {\rm{s.t.}}\ {n_{l'}} - {n_l} = i,\\
 &{\bf{0}},\ \ \ \ {\rm{otherwise}}.
 \end{split} \right.
 \end{equation}
 \begin{figure}[!t]
 \setlength{\abovecaptionskip}{-0.1cm}
 \setlength{\belowcaptionskip}{-0.1cm}
 \centering
 \centerline{\includegraphics[width=3.5in,height=1.05in]{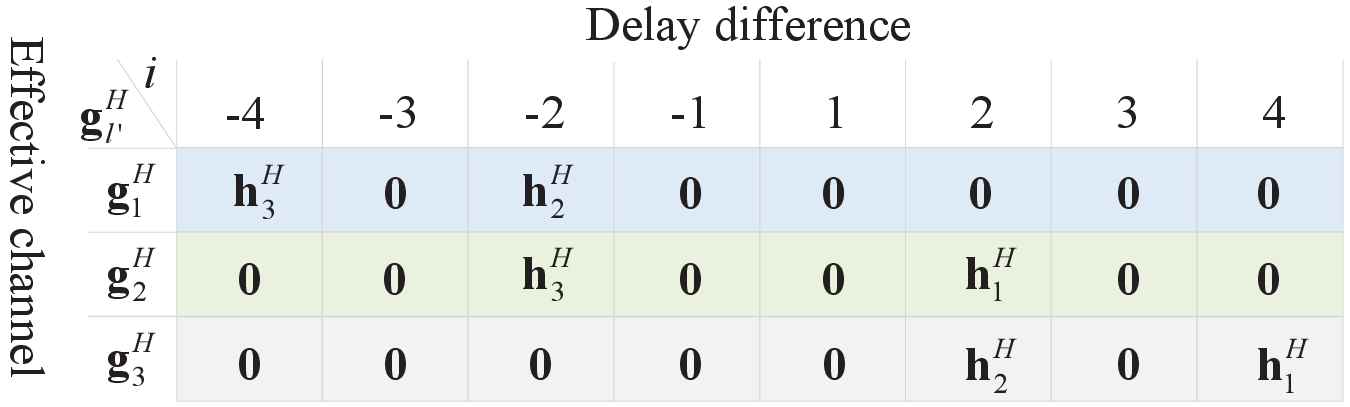}}
 \caption{An illustration of the effective channel with $L =3$ multi-paths, with $n_1=1$, $n_2=3$, and $n_3=5$. The delay spread is ${n_{{\rm{span}}}} = {n_3} - {n_1} = 4$, and the delay difference $i \in \left\{ { \pm 1, \pm 2, \pm 3, \pm 4} \right\}$. For example, $n_1 - n_2= -2$, ${\bf{g}}_1^H\left[ { - 2} \right] = {\bf{h}}_2^H$, and $n_1 - n_3= -4$, ${\bf{g}}_1^H\left[ { - 4} \right] = {\bf{h}}_3^H$.}
 \label{delayDifferenceillustration}
 \vspace{-0.7cm}
 \end{figure}

 A simple illustration with $L=3$ multi-paths is given in Fig.~\ref{delayDifferenceillustration}. It is worth mentioning that for different multi-path pairs $\left( {l,l'} \right)$ in \eqref{delayDifferenceCompensationReceivedSignal}, their delay difference ${\Delta _{l',l}}$ might be identical, as shown in Fig.~\ref{delayDifferenceillustration} for $i=-2$. Therefore, their corresponding interfering symbols need to be grouped. To this end, \eqref{delayDifferenceCompensationReceivedSignal} is equivalently written as
 \begin{equation}\label{delayDifferenceCompensationEquivalentReceivedSignal}
 \hspace{-0.5ex}
 \begin{aligned}
 &y\left[ n \right]= \Big( {\sum\nolimits_{l = 1}^L {{\bf{h}}_l^H{{\bf{f}}_l}} } \Big)s\left[ {n - {n_{\max }}} \right] + \\
 &\sum\nolimits_{i =  - {n_{{\rm{span}}}},i \ne 0}^{{n_{{\rm{span}}}}} {\Big( {\sum\nolimits_{l' = 1}^L {{\bf{g}}_{l'}^H\left[ i \right]{{\bf{f}}_{l'}}} } \Big)s\left[ {n - {n_{\max }} + i} \right]}  + z\left[ n \right].
 \end{aligned}
 \end{equation}
 Since $s\left[ n \right]$ is independent across $n$, the resulting SINR is
 \begin{equation}\label{delayDifferenceSINRForm1}
 \begin{aligned}
 {\gamma} &= \frac{{{{\Big| {\sum\nolimits_{l = 1}^L {{\bf{h}}_l^H{{\bf{f}}_l}} } \Big|}^2}}}{{\sum\nolimits_{i =  - {n_{{\rm{span}}}},i \ne 0}^{{n_{{\rm{span}}}}} {{{\Big| {\sum\nolimits_{l' = 1}^L {{\bf{g}}_{l'}^H\left[ i \right]{{\bf{f}}_{l'}}} } \Big|}^2}}  + {\sigma ^2}}}\\
 &= \frac{{{{{\bf{\bar f}}}^H}{\bf{\bar h}}{{{\bf{\bar h}}}^H}{\bf{\bar f}}}}{{{{{\bf{\bar f}}}^H}\left( {\sum\nolimits_{i =  - {n_{{\rm{span}}}},i \ne 0}^{{n_{{\rm{span}}}}} {{\bf{\bar g}}\left[ i \right]{{{\bf{\bar g}}}^H}\left[ i \right]}  + {\sigma ^2}{\bf{I}}/{{\left\| {{\bf{\bar f}}} \right\|}^2}} \right){\bf{\bar f}}}},
 \end{aligned}
 \end{equation}
 where ${\bf{\bar h}} = {\left[ {{\bf{h}}_1^T, \cdots ,{\bf{h}}_L^T} \right]^T} \in {{\mathbb{C}}^{ML \times 1}}$, ${\bf{\bar f}} = {\left[ {{\bf{f}}_1^T, \cdots ,{\bf{f}}_L^T} \right]^T}$ $\in {{\mathbb{C}}^{ML \times 1}}$ and ${\bf{\bar g}}\left[ i \right] = {\left[ {{\bf{g}}_1^T\left[ i \right], \cdots ,{\bf{g}}_L^T\left[ i \right]} \right]^T} \in {{\mathbb{C}}^{ML \times 1}}$.

 If the low-complexity ISI-MRT beamforming is applied, i.e., ${\bf{\bar f}} = \sqrt P {\bf{\bar h}}/{{\left\| {{\bf{\bar h}}} \right\|}}$, the received SINR can be expressed as
 \begin{equation}\label{residualISIMRTBeamforming}
 {\gamma _{{\rm{MRT}}}} = \frac{{P{{\left\| {{\bf{\bar h}}} \right\|}^2}}}{{P\sum\nolimits_{i =  - {n_{{\rm{span}}}},i \ne 0}^{{n_{{\rm{span}}}}} {{{\left| {{{{\bf{\bar g}}}^H}\left[ i \right]{\bf{\bar h}}/\left\| {{\bf{\bar h}}} \right\|} \right|}^2}}  + {\sigma ^2}}}.
 \end{equation}
 In particular, when the channel is sparse and/or the antenna dimension is large so that all $L$ multi-paths are well separated at the transmitter, we have ${{\bf{\bar g}}^H}\left[ i \right]{\bf{\bar h}}/\left\| {{\bf{\bar h}}} \right\| \to 0$, and \eqref{residualISIMRTBeamforming} reduces to
 \begin{equation}\label{reducedResidualISIMRTBeamforming}
 {\gamma _{{\rm{MRT}}}} \to \bar P{\left\| {{\bf{\bar h}}} \right\|^2} = \bar P\sum\nolimits_{l = 1}^L {{{\left\| {{{\bf{h}}_l}} \right\|}^2}},
 \end{equation}
 which is in accordance with \eqref{DAMReceivedSNRMRTExpression}.

 To obtain the optimal beamforming design for DAM, it is noted that the SINR in \eqref{delayDifferenceSINRForm1} is a generalized Rayleigh quotient with respect to ${{\bf{\bar f}}}$, which is maximized by the MMSE beamforming
 \begin{equation}\label{beamformingForResidualISI}
 {{\bf{\bar f}}^{{\rm{MMSE}}}} = \sqrt P {{{\bf{C}}^{ - 1}}{\bf{\bar h}}}/{\left\| {{{\bf{C}}^{ - 1}}{\bf{\bar h}}} \right\|},
 \end{equation}
 where ${\bf{C}}{\rm{ }} \buildrel \Delta \over = \sum\nolimits_{i =  - {n_{{\rm{span}}}},i \ne 0}^{{n_{{\rm{span}}}}} {{\bf{\bar g}}\left[ i \right]{{{\bf{\bar g}}}^H}\left[ i \right]}  + \frac{{{\sigma ^2}}}{P}{\bf{I}}$ denotes the interference-plus-noise covariance matrix. The resulting SINR is
 \begin{equation}\label{MMSEOptimalSINR}
 {\gamma _{{\rm{MMSE}}}} = {{{\bf{\bar h}}}^H}{{\bf{C}}^{ - 1}}{\bf{\bar h}}.
 \end{equation}
\vspace{-0.7cm}
\subsection{DAM Versus OFDM}
\begin{figure}[!t]
  \setlength{\abovecaptionskip}{-0.1cm}
  \setlength{\belowcaptionskip}{-0.1cm}
  \centering
 \centerline{\includegraphics[width=3.1in,height=0.85in]{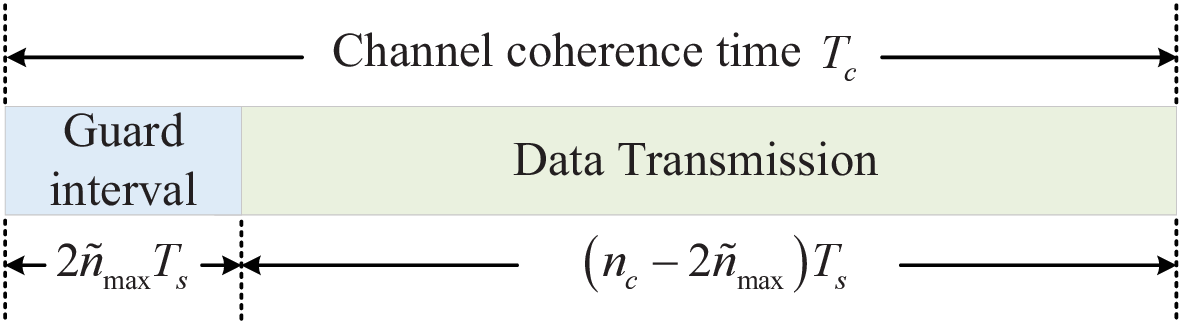}}
  \caption{An illustration of DAM block structures, where DAM only requires a guard interval for each channel coherence block.}
  \label{DAMOFDMBlockStructure}
  \vspace{-0.65cm}
 \end{figure}

In this subsection, we compare the performance between DAM and OFDM. Denote by ${T_c}$ the channel coherence time, within which the channel is assumed to be unchanged. Let ${n_c} \approx {T_c}/{T_s}$ denote the number of single-carrier symbols within each channel coherence block, where ${T_s} = 1/B$. Further denote by ${\tilde n_{\max }}$ an upper bound of the maximum delay over all channel coherence blocks, i.e., ${{\tilde n}_{\max }} \ge {n_{\max }}$. For the proposed DAM, It is observed from \eqref{DelayCompensationReceivedSignal} that to avoid ISI across different channel coherence blocks, a guard interval of length $2{\tilde n_{\max }}$ is needed for each coherence block, as shown in Fig.~\ref{DAMOFDMBlockStructure}. Thus, the effective spectral efficiency of DAM with ISI-ZF, ISI-MRT and ISI-MMSE beamforming schemes in bits/second/Hz (bps/Hz) can be written as
\begin{equation}\label{DAMCapacity}
{C_{{\rm{DAM}}}} = \frac{{\left( {{n_c} - 2{{\tilde n}_{\max }}} \right)}}{{{n_c}}}{\log _2}\left( {1 + {\gamma _a}} \right),
\end{equation}
where $a \in \left\{ {{\rm{ZF,MRT,MMSE}}} \right\}$, and the guard interval overhead of DAM is $2{{\tilde n}_{\max }}/{n_c}$.

On the other hand, for OFDM with $K$ sub-carriers, a cyclic prefix (CP) of length ${\tilde n}_{\max }$ needs to be inserted for each OFDM symbol, which has duration $\left( K + {{\tilde n}_{\max }} \right){T_s}$. The number of OFDM symbols for each channel coherence time is ${n_{{\rm{OFDM}}}} = {n_c}/\left( {K + {{\tilde n}_{\max }}} \right)$. Therefore, by applying the optimal MRT beamforming for each of the $K$ sub-carriers, the effective spectral efficiency of OFDM is
\begin{equation}\label{OFDMCapacity}
{C_{{\rm{OFDM}}}}{\rm =} \frac{{{n_c} - {n_{{\rm{OFDM}}}}{{\tilde n}_{\max }}}}{{{n_c}}}\frac{1}{K}\sum\nolimits_{k = 1}^K {{{\log }_2}\Big( {1 + \frac{{{p_k}{{\left\| {{\bf{h}}\left[ k \right]} \right\|}^2}}}{{\sigma ^2}/K}} \Big)},
\end{equation}
where $p_k$ and ${\bf{h}}\left[ k \right]$ denote the transmit power and the frequency-domain channel of the $k$th sub-carrier, respectively, and the optimal $p_k$ can be obtained by the classic water-filling (WF) power allocation \cite{goldsmith2005wireless}. By comparing \eqref{DAMCapacity} and \eqref{OFDMCapacity}, it is observed that DAM not just avoids the practical issues of OFDM like high PAPR and severe OOB emission as well as vulnerability to CFO with low-complexity single-carrier transmission, but also significantly reduces the guard interval overhead since ${n_{{\rm{OFDM}}}} \gg 1$ typically holds.

%

\vspace{-0.3cm}
\section{Simulation Results}\label{sectionNumericalResults}
In this section, simulation results are provided to evaluate the performance of the proposed DAM technique. We assume that the carrier frequency is $f = 28$ GHz, the total bandwidth is $B=128$ MHz, the noise power is ${\sigma ^2} =  - 85$ dBm, and the transmit power is $P=30$ dBm. The transmitter is equipped with an uniform linear array (ULA) with adjacent elements separated by half-wavelength. The channel coherence time is ${T_c} = 1$ ms, and the total number of single-carrier symbols within each coherence time is thus ${n_c} = 1.28 \times {10^5}$. Note that for mmWave communications, the number of multi-paths is usually small since the high frequency signals suffer from severe scattering and diffraction losses \cite{akdeniz2014millimeter,zeng2016millimeter}. Thus, unless otherwise stated, the number of temporal-resolvable multi-paths is set as $L=5$, which are uniformly distributed in $\left[ {0,{\tau_{\rm{max}}}} \right]$, with $\tau_{\rm{max}} = 312.5$ ns. The number of sub-paths $\mu_l$, $\forall l$ is uniformly distributed in $\left[ {1,{\mu_{\rm{max}}}} \right]$, with ${\mu_{\max}}=3$. The AoDs of all the sub-paths are randomly distributed in the interval $\left[ { - {{60}^\circ },{{60}^\circ }} \right]$. Furthermore, the complex-valued gains ${\alpha _l}$, $\forall l$, are generated based on the model developed in \cite{akdeniz2014millimeter}. For the benchmark of OFDM scheme, the number of sub-carriers is $K=512$, and a CP of length ${\tilde n}_{\max } = 40$ is used. The number of OFDM symbols within the coherence time is then given by ${n_{{\rm{OFDM}}}} = \frac{{{T_c}}}{{\left( K + {{{\tilde n}_{\max }} } \right){T_s}}} \approx 231$.

 \begin{figure}[!t]
  \setlength{\abovecaptionskip}{-0.18cm}
  \setlength{\belowcaptionskip}{-0.1cm}
  \centering
  \centerline{\includegraphics[width=3.5in,height=2.6in]{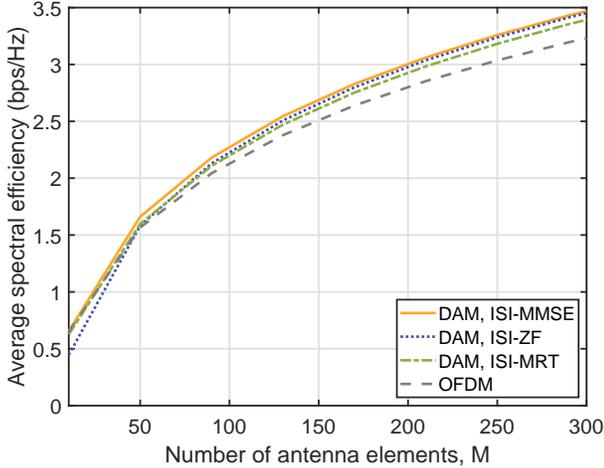}}
  \caption{Average spectral efficiency versus the antenna number for OFDM and DAM with ISI-MMSE, ISI-ZF, and ISI-MRT beamforming.}
  \label{MMSEZFMRTCapacityVersusAntennaNumber}
  \vspace{-0.5cm}
 \end{figure}

Fig.~\ref{MMSEZFMRTCapacityVersusAntennaNumber} shows the average spectral efficiency over ${10^4}$ channel realizations versus the antenna number $M$ for DAM with ISI-MMSE, ISI-ZF, and ISI-MRT beamforming, together with the benchmark of OFDM scheme. It is observed that all the three beamforming schemes for the proposed DAM give similar performance. This is expected since when the number of antennas $M$ is much larger than that of multi-path delays $L$, the multi-path signal components can be well separated in space for all the three beamforming schemes, which is in accordance with \eqref{DAMReceivedSNRMRTExpression} and \eqref{reducedResidualISIMRTBeamforming}. It is also observed from Fig.~\ref{MMSEZFMRTCapacityVersusAntennaNumber} that the proposed DAM technique, though with simpler complexity, still outperforms the classic OFDM, thanks to the saving of guard interval overhead, as illustrated in Fig.~\ref{DAMOFDMBlockStructure}. Specifically, for the considered setup, the overhead for DAM is $2{{\tilde n}_{\max }}/{n_c} = \frac{{80}}{{1.28 \times {{10}^5}}} = 0.0625\% $, and that for OFDM is ${n_{\rm{OFDM}}}{{\tilde n}_{\max }}/{n_c} = \frac{{231 \times 40}}{{1.28 \times {{10}^5}}} = 7.2\% $.

%

 \begin{figure}[!t]
  \setlength{\abovecaptionskip}{-0.18cm}
  \setlength{\belowcaptionskip}{-0.1cm}
  \centering
  \centerline{\includegraphics[width=3.5in,height=2.6in]{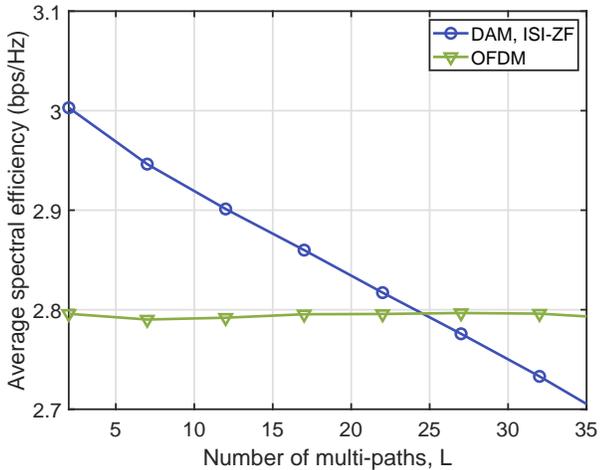}}
  \caption{Average spectral efficiency versus the number of temporal-resolvable multi-paths for DAM and OFDM.}
  \label{OFDMDAMSEVersusDelayNumber}
  \vspace{-0.7cm}
 \end{figure}

 Fig.~\ref{OFDMDAMSEVersusDelayNumber} shows the average spectral efficiency versus the number of temporal-resolvable multi-paths $L$. The number of antennas is $M=200$. It is observed that the average spectral efficiency of OFDM remains almost unchanged as $L$ increases, while that of DAM gives the best performance when $L$ is small. This is expected since delay alignment is most effective when the number of antennas $M$ is considerably larger than that of multi-paths $L$, so that there is little loss in spatial beamforming gain after delay alignment. It is observed from Fig.~\ref{OFDMDAMSEVersusDelayNumber} that for the considered setup, DAM outperforms OFDM for up to $L=25$ multi-paths, and this number is expected to be even larger when $M$ further increases.

\vspace{-0.4cm}
\section{Conclusion}
This paper proposed a novel broadband transmission technology termed DAM, by exploiting the abundant spatial dimension of large antenna arrays and the multi-path sparsity of mmWave and Terahertz channels. DAM enabled a low-complexity and equalization-free single-carrier communication, yet without suffering from the ISI issue. The proposed DAM was studied for a MISO system, where ISI-ZF, ISI-MRT, and ISI-MMSE beamforming were proposed. Simulation results demonstrated the superior performance of the proposed DAM technique than the classical OFDM, in terms of the spectral efficiency and signal processing complexity.

\vspace{-0.5cm}



\bibliographystyle{IEEEtran}
\bibliography{refDAM}
\end{document}